\documentclass[12pt]{article}
\usepackage{setspace,psfig,graphicx,subfigure,jmb}
\usepackage{overcite}

\makeatletter
   \def\@cite#1{\textsuperscript{#1}}
 \makeatother

\begin{document}

\doublespacing

\title{Understanding the determinants of stability and folding of small globular proteins from their energetics}

\author{G. Tiana$^{1,2}$, F. Simona$^1$, Giacomo M.S. De Mori$^3$, \\
R. A. Broglia$^{1,2,4}$, G. Colombo$^{3,*}$ \\
$^1$ Department of Physics, University of Milano, \\via Celoria 16, 20133 Milano, Italy\\
$^2$ INFN, sez. di Milano via Celoria 16, 20133 Milano, Italy\\
$^3$ Istituto di Chimica del Riconoscimento Molecolare, CNR \\via Mario Bianco 9, 20131 Milano, Italy\\
$^4$ The Niels Bohr Institute, Bledgamsvej 17, \\2100 Copenhagen, Denmark}

\date{\today}
\maketitle

{*} Author to whom correspondence should be addressed: g.colombo@icrm.cnr.it

\newpage

\begin{abstract}
The results of minimal model calculations suggest that the stability and the kinetic accessibility of the native state of small globular proteins are controlled by few "hot" sites. By mean of molecular dynamics simulations around the native conformation, which simulate the protein and the surrounding solvent at full--atom level, we generate an energetic map of the equilibrium state of the protein and simplify it with an Eigenvalue decomposition. The components of the Eigenvector associated with the lowest Eigenvalue indicate which are the "hot" sites responsible for the stability and for the fast folding of the protein. Comparison of these predictions with the results of mutatgenesis experiments, performed for five small proteins, provide an excellent agreement. 
\end{abstract}


\newpage

\section{Introduction}

In the last few years, the study of how the structure and stability of a protein is connected with its sequence has been the major focus of research of a wide number of scientists. In particular, the problems of protein folding and of the structure--stability relationship have been tackled through the application of a diverse set of experimental and theoretical techniques. All these studies have highlighted the role of the free energy landscape of proteins to understand the properties of the associated sequences. The discussion has been turned then to the problem of determining operatively the free energy landscape of a specific protein. 

Experimental techinques have yielded detailed information on macroscopic features of protein dynamical beahviour, such as folding times and stability, and on some specific issues at the level of amino acids, such as the sensitivity to mutations (cf. Fersht, 1999\cite{fersht}). However, there is still no experimental procedure capable of providing an insights at amino acid level into either the folding process in its completeness or the stabilization determinants of proteins. In order to obtain such a detailed description, one must turn to theoretical and computational methods.   

Realistic models can be used to give an exaustive description of the free energy landscape only to short peptides, due to the high computational cost involved. Ferrara and Caflish, for instance, could reconstruct the whole free energy landscape of a small designed \(\beta \)-sheet peptide with an all--atom representation. Daura \emph{et al.} were able to demonstrate the reversible folding of a small (seven residues) helix forming \(\beta \)-peptide in methanol solvent using long molecular dynamics (MD) simulations. 

Minimal models \cite{sh1,dill} have provided interesting results about the general features of the free energy landscape of proteins. They give an approximate description of both the interaction energy among amino acids (usually through a contact potential encoded in a 20$\times$20 matrix, e.g. Miyazawa, 1985\cite{mj}), and of the entropy (the protein being described as a chain of beads on the verteces of a cubic lattice). Making use of this kind of model,  it has been possible to understand the basic energetic properties which characterize a folding sequence, that is a large gap between the energy of the native and of the lowest, structurally dissimilar state \cite{sh1}. Moreover, it has been suggested that the process of folding takes place in a hierarchical fashion, being guided by the formation of local elementary structures, that is short segments of the protein stabilized by strong interactions, which assemble together into a nucleus, which sits in the energetic basin of attraction of the native state \cite{jchemphys2,jchemphys3}. Such hierarchical mechanism makes the folding process fast by gradually squeezing out the entropy from the system. Although very interesting from the point of view of general principles of folding, this approach is not very helpful while dealing with real, well-defined proteins.

Other approachs focus on the geometric features of the protein, providing a detailed picture of the structure and using simple approximations for the potential function. From the physical point of view, this means focusing the attention to the entropic part of the energy landscape, neglecting the energetics. An example of this approach are the works by Baker \cite{baker}, who correlates the folding rate of a number of proteins with a "contact order" parameter, which accounts for the locality of the interactions. The Go model is another approximation which describes well the entropy of the protein chain, making crude approximation on the interaction energy \cite{go} (the energy is a contact function which assumes the value $-1$ if the contact is a native contact and zero otherwise).

In the present work we use a novel simple approach to extract information on the folding process from the structure of the native state, based on a detailed description of the interaction energy and neglecting the entropy of the protein. The rational behind this choice is that the stabilization energy is distibuted quite unevenly in proteins, as testified by the fact that mutations in most of the sites of a protein have little effect on its folding properties, while there are few key sites, where the stabilization energy is concentrated, which are highly sensitive to mutations \cite{jchemphys1}. These sites are those which, in the analysis performed with minimal models, build out the local elementary structures which controls the folding process \cite{jchemphys2}. Consequently, from the localization of these sites one can not only understand the mechanism which stabilizes the protein, but also get a qualitative insight in the kinetics mechanism of folding.

The understanding of how the different amino acids determine the free energy of the protein, and expecially of what are the residues which play a key role in its the stability and kinetics, 
is of great interest not only in the understanding of how the one--dimensional protein sequence encodes for the three--dimensional native conformation. In fact, this could help in the design of new proteins with specific tasks (see, e.g. Dahiyat,1997\cite{Dahiyat97}) or to manipulate existing proteins \cite{Colombo99}.

In Section: "The method and its validation from a minimal model" we describe the tool used for detecting the key sites and test the method on a lattice model protein, whose kinetics and thermodynamics are perfectly under control. In Section: "Application to proteins" we repeat the procedure, making use of all--atom molecular dynamics simulations, on five small globular proteins, and compare the sites identified by our procedure with those indicated by experiments as crucial for the folding mechanism and for protein stabilization. The last two sections  contain the discussion of the results and the conclusions.

\section{The method and its validation from a minimal model}

The basic idea of the present analysis is to extract energetic information on the protein from molecular dynamics (MD) simulations, and from here to get insights into the determinants of the stability of the native protein conformation, and their influence on  the folding process. The main tool to achieve this goal is the amino acid--amino acid interaction matrix $M_{ij}$, calculated averaging each amino acid--amino acid interaction energy, comprising all the non-bonded inter-residue energy components (e.g. van der Waals and Electrostatic), over a MD trajectory starting from the native conformation. The matrix $M_{ij}$ can be decomposed in Eigenvalues, in the form
\begin{equation}
\label{decomp}
M_{ij}=\sum_{\alpha=1}^N \lambda_\alpha \mu_i^\alpha \mu_j^\alpha,
\end{equation}
where $N$ is the number of amino acids in the protein, $\lambda_\alpha$ is an Eigenvalue and $\mu_i^\alpha$ are the components of the associated Eigenvector. We assume that the Eigenvectors are normalized to unity and, since $M_{ij}$ is symmetrical, all the Eigenvalues are real. 

For sake of simplicity, we label the $N$ Eigenvalues in increasing order, so that $\lambda_1$ is the most negative. Accordingly, the different terms in the sum (\ref{decomp}) approximate the real interaction energy $M_{ij}$ to an increasing extent, the first term containing the largest contribution to the stabilization of the native conformation (to the order of magnitude of $\lambda_1$). The components $\mu^1_i$ of the associated Eigenvector indicate to which extent each amino acid participates to the stabilization. In other words, each term in (\ref{decomp}) accounts for an amount of energy $\lambda_\alpha$ which is shared among the different residues according to the corresponding Eigenvector $\mu^\alpha_i$.

If the second Eigenvalue $\lambda_2$ is much higher than $\lambda_1$, one can approximate the whole interaction matrix as 
\begin{equation}
\label{approx}
M_{ij}=\lambda_1\mu^1_i\mu^1_j,
\end{equation}
reducing the information needed to specify the interaction from $N^2$ to $N$ numbers.

This approximation reduces a complicated two--body interaction into a kind of interaction determined by "charges" (as the electrostatic interaction). It states that there are some amino acids which are strongly interacting and others which are weakly interacting {\it tout court}, depending on the "charges" $\mu^1_i$ and $\mu^1_j$ of the two amino acids. Consequently, if amino acid A attracts strongly amino acids B (e.g., both with large, positive "charge"\footnote{To be noted that, unlike the case of electrostatic interaction, here "charges" with the same sign attract each others, while "charges" with opposite sign repell each others. This is because $\lambda_1$ in Eq. (\ref{approx}) is negative, being defined as the most negative of the Eigenvalues.}), and amino acid B attracts strongly amino acid C (e.g., both with positive, large charge), in the present approximation A attracts strongly C. The opposite is also true: if the interactions A--B and B--C are weak, also the interaction A--C will be weak. On the other hand, the real interaction $M_{ij}$ could be more complicated than this, being able to account for any combination of attractive and repelling pairs (e.g., A attracts B, B attracts C, but A repels C).

Consequently, that of Eq (\ref{approx}) is, in priciple, a very crude approximation.
On the other hand, from the analysis of minimal models it emerged that proteins are composed of a nucleus, which contains most of the stabilization energy and which displays a rich network of interaction \cite{nucleus,jchemphys2}. This network of interaction makes each residue of the nucleus interact favorably with many other residues of the nucleus. A nucleus where each residue attracts each other residue of the nucleus can be easily described by "charges" (i.e., the residues belonging to the nucleus display a large "charge", while the other residues a much lower "charge").
Consequently, if the energetics of the protein can be descibed by "charges", the approximation (\ref{approx}) results accurate.

A method to assess precisely the quality of the approximation is to study the Eigenvalues spectrum. If $\lambda_1$ is much smaller (i.e., more negative) than the other Eigenvalues, it means that the approximation is good. 

To be noted that this method, although being an application of the well--known principal component analysis (PCA), has nothing to do with the various kinds of conformational PCA, or "essential dynamics", present in the literature \cite{Amadei93}, whose goal is to effectively decrease the dimensionality of the conformational space. 

\begin{figure}
\centerline{\psfig{file=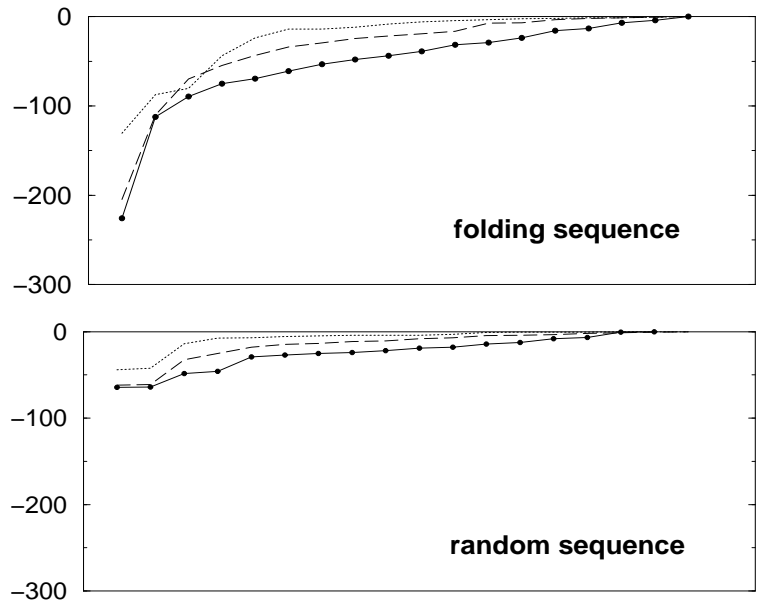,width=10cm,height=8cm,angle=-90}}
\caption{The negative Eigenvalues of the interaction matrix in the lattice model. The Eigenvalues are plotted in ascending order, for a folding sequence and for a random, non--folding sequence. The solid curve is associated with the interaction in the native conformation. The dashed curve displays the Eigenvalues of the average interaction matrix, calculated starting from the native conformation and is allowed to fluctuate for $10^4$ steps at the folding temperature, while the dotted curve is associated with the average interaction starting from a random conformation and letting the system evolve for $10^4$ steps.  }
\label{EV}
\end{figure}

To test the above procedure, we have studied a 36mer lattice model sequence designed to fold to a unique conformation, whose nucleus is known to be built out of the three local elementary structures 3--6, 11--14 and 27--30 \cite{jchemphys2}. The spectrum of Eigenvalues of the interaction matrix calculated in the native conformation is displayed as a solid curve in the upper part of Fig. \ref{EV}. For comparison, the spectrum associated with a random sequence is displayed in the lower part of the same figure. It is clear that, in the case of the designed sequence, the lowest Eigenvalue is well separated from the others ($\lambda_2-\lambda_1=113.6$, where the average spacing between the others is $13.3$), while in the case of the random sequence this is not true ($\lambda_2-\lambda_1=0.4$, the average spacing between the others being $4.0$). As discussed above, this difference reflects the fact that the designed sequence displays a nucleus of mutually interacting residues, where the stabilization energy is concentrated, while in the random sequence there is a complicated pattern of attractive and repulsive contacts.  

Since for real proteins the equilibrium state, at the length scale of single amino acids, is an ensemble of different conformations \cite{frauenfelder} sharing the same overall topology, we have repeated the calculation letting the system fluctuate among these conformations. This is done by performing a Monte Carlo search ($10^4$ MC steps) at folding temperature and calculating the interaction energy between pairs of amino acids averaged over these Monte Carlo steps. The resulting Eigenvalue spectrum, displayed as a dashed curve in Fig. \ref{EV}, is qualitatively equal to that calculated in the unique native conformation. 

\begin{figure}
\centerline{\psfig{file=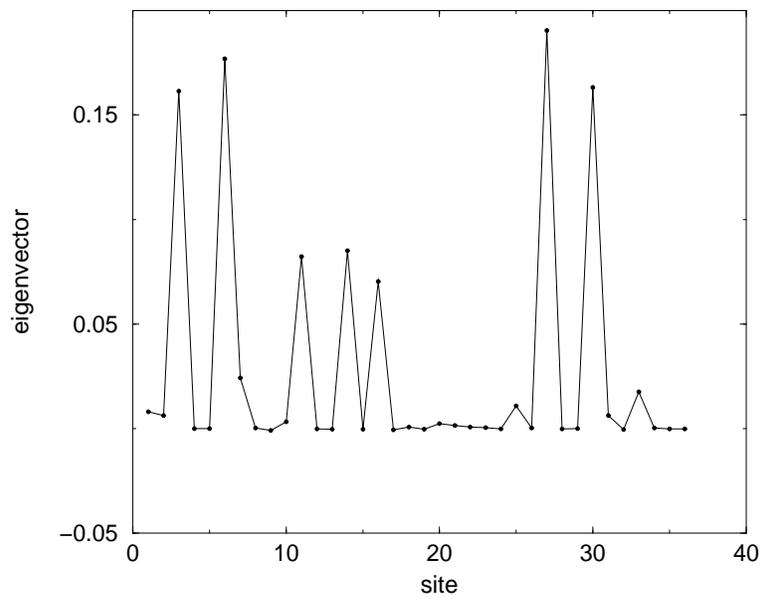,width=10cm,height=8cm,angle=-90}}
\caption{The Eigenvector associated with the largest Eigenvalue for the lattice--model folding sequence (corresponding to the first point of the dashed curve in the upper part of Fig. \protect\ref{EV}.}
\label{EVV}
\end{figure}

A difference between the Eigenvalue spectra of the folding and of the random sequence is detectable also for interaction matrices calculated starting from random conformations and performing the average over a time which is much smaller than the overall folding time ($10^3$ MC steps, compared to a characteristic folding time of $8\cdot 10^5$ MC steps for the folding sequence). The results are displayed as a dotted curve in Fig. \ref{EV}. For the folding sequence $\lambda_2-\lambda_1=43.2$, to be compared to the average spacing of $7.6$, while for the random sequence $\lambda_2-\lambda_1=1.1$ and the average spacing is $2.6$. The reason for this behavior is that local elementary structures, which are part of the nucleus and carry some of the stabilization energy of the protein, are formed at the very early stages of the folding process of good sequences. On the other hand, random sequences display a disordered collapse not displaying any internal structure. 
The details about how the stabilization energy is distributed among the amino acids are contained in the normalized Eigenvector $\mu^1_i$ associated to the lowest Eigenvalue. In Fig. \ref{EVV} are displayed the components of such an Eigenvector in the case of the interaction matrix calculated in the native conformation of the designed sequence (corresponding to the solid curve in the upper part of Fig. \ref{EV}). The plots corresponding to the other two curves in the upper part of Fig. \ref{EV} are similar. It is easy to appreciate the fact that the plot displays six peaks corresponding to the amino acids which build out the local elementary structures (residues 3,6,11,14,27 and 30) and one (residue 16) that, although not participating to a local elementary structure, interacts strongly with the nucleus. Anyway, these seven sites are exactly those that, if mutated, lead to the denaturation of the protein, and are called "hot" and "warm" sites in the paper by Tiana {\it et al.}\cite{jchemphys1}.

The remarkable aspect of obtaining information about the stability of the folded state and the folding process from the native conformation, from fluctuations about it and from the early stages of folding is that these calculations can be easily repeated for real proteins in the framework of a more realistic, all--atom model also including the solvent. 

\section{Application to proteins}

Five different and experimentally well--characterized proteins were used to test the model with realistic all--atom simulations including an atomic representation of the solvent:  the \( \alpha  \)-spectrin SH3 domain, the src SH3 domain,  the IgG binding domain of protein G (herein called simply Protein G), the IgG binding domain of protein L (shortly called protein L) and Chymotrypsin Inhibitor 2 (CI2). For the sake of simplicity in the discussion of the eigenvalue and of the eigenvector properties, the numbering followed here for each protein starts from residue number 1, and the correspondent numbering used in the pdb and in other papers will be reported every time in parentheses.

All of these systems have been studied in depth in terms of the mutations needed to (de)stabilize their folded states and to influence their folding kinetics. Most of these works were based on a massive experimental analysis performed through point mutations and on the characterization of the effects of every single mutation on the thermodynamical stability and on the kinetical properties of the protein. The ability to identify the mutation sites via a fast and simple computational method  becomes thus an extremely appealing feature of all-atom molecular simulations. The five systems were analyzed through 10 ns long all-atom molecular dynamics (MD) simulations, using an explicit representation of the solvent and the Particle Mesh Ewald method to calculate electrostatic interactions, in order to avoid cut-off induced artifacts.

\begin{figure}
\centerline{\psfig{file=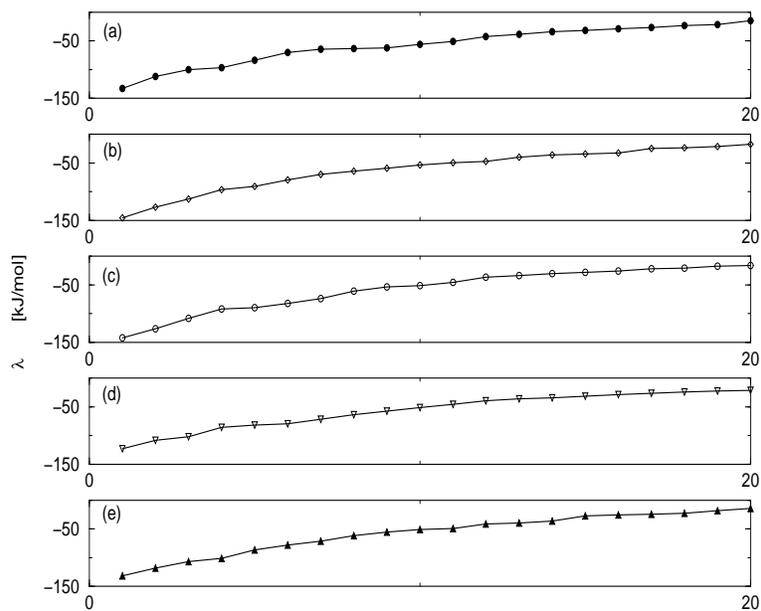,width=10cm,height=8cm,angle=-90}}
\caption{The negative Eigenvalues of the interaction matrix for the all-atom MD simulations. The Eigenvalues are plotted in ascending order:  (a) filled circles refer to the \( \alpha  \)-spectrin SH3 domain, (b) diamonds to the src SH3 domain, (c) open circles refer to the IgG binding domain of protein G, (d) open  triangles refer to the IgG binding domain of protein L and (e) filled triangles refer to CI2.}
\label{ev_md}
\end{figure}

The lower part of the Eigenvalue spectrum, consisting of the first twenty eigenvalues for each of the five proteins are displayed in Fig. \ref{ev_md}(a-e). 
In all the proteins examined the separation $\Delta\lambda$ between eigenvalues $\lambda_1$ and $\lambda_2$ is much larger than the average spacing $\overline{\lambda}$ between the other Eigenvalues (cf. Table 1), as in the case of the lattice model protein discussed above. Only in the case of CI2, this property becomes marginal and, in particular, the first three Eigenvalues are almost equally spaced. The Eigenvector associated with the first Eigenvalue can thus be approximated to contain the most relevant energetic information of the five proteins studied. 

One has anyway to notice that the ratio between $\Delta\lambda$ and $\overline{\lambda}$ is somewhat smaller for the five real proteins than for the idealized lattice protein. There are three reasons contributing to this fact: 1) we use a general--purpose force field, which has not been optimally designed for the particular fold of the proteins under study, 2) the noise present in MD simulations due to the numerical approximations in the integration steps can flatten the Eigenvalue spectra of the five proteins, 3) real proteins can concentrate energy in clusters of amino acids for reasons different than their folding properties, such as for binding other molecules, for enzymatic activity, etc.

\subsection{$\alpha$ spectrin SH3 domain}

The spectrum of eigenvalues for $\alpha$--spectrin SH3 domain is reported in Fig. \ref{ev_md}(a) (filled circles). In this case, the difference between the first two lowest eigenvectors is $\Delta\lambda=20.9$, which means five times the average spacing $\overline{\lambda}=4.2$, allowing the approximation of Eq. \ref{approx}.

\begin{figure}
\centerline{\psfig{file=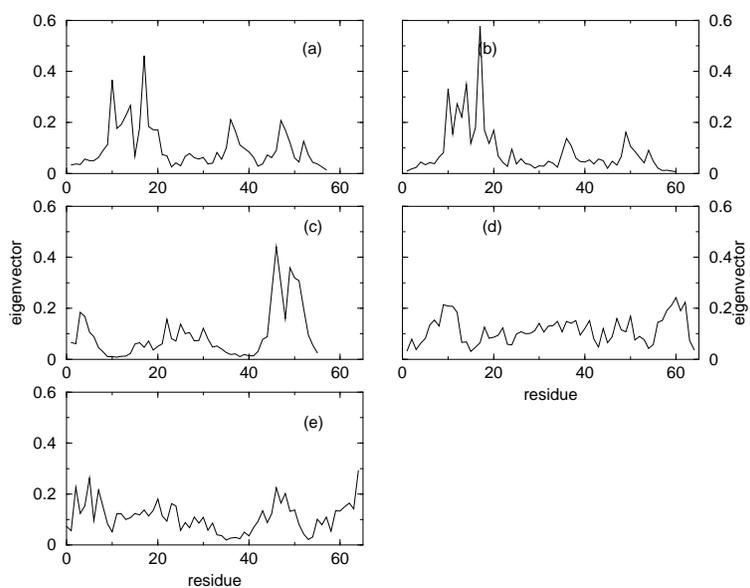,width=10cm,height=8cm,angle=-90}}
\caption{The Eigenvectors associated with the largest Eigenvalue in the all-atom MD simulations:  (a) \( \alpha  \)-spectrin SH3 domain, (b)  src SH3 domain, (c) IgG binding domain of protein G, (d) IgG binding domain of protein L and (e) CI2.}
\label{vec_md}
\end{figure}

In Fig. \ref{vec_md}(a) are displayed the components $\mu^1_i$ of the normalized Eigenvector associated with the lowest Eigenvalue, components which indicate how the stabilization energy is distributed (under the approximation of Eq. \ref{approx}) among the aminoacids of the $\alpha$--spectrin SH3 domain. The plot displays particularly large values of the components in the intervals 10--20 (15--25 in the original numbering by Serrano and coworkers), 34--39 (39-44) and 46--49 (51-54) (in particular, the maximum of the peaks are at sites 10, 18, 36, 47 and 48).

This protein has been extensively studied by Martinez, Pisabarro and Serrano \cite{serrano}. By means of point mutations they showed that residues V18 (V23), V39 (V44) and V48 (V53) are important for the stability of the protein, as their mutation leads to a strong destabilization of the native state (cf. Table 2). These sites are indeed among those showing, in our analysis, a large value of the Eigenvector component. This is a straighforward result, considering that the components of the Eigenvector express the degree to which a given residue partecipate to the stability of the protein.

A more striking result is that also the residues which play a key role in the {\it kinetics} display large values of the Eigenvector components. In fact, Martinez and coworkers showed that the formation of the distal hairpin (residues 38--48 (43--53)) and the anchoring of the second strand of the RT loop (residues 18--19 (24--24)) are determinant in the kinetics of folding (cf. Martinez\cite{serrano} and Table 2). The highest peak in Fig. \ref{vec_md}(a) corresponds to residue V18 (RT loop), while the two peaks centered at residues W36 (W41) and V48 (V53) correspond to the sites stabilizing the distal hairpin.

The tight connection between the energetic properties of a residue and its role in the folding kinetics is not unexpected and can be easily rationalized in terms of stabilization of local elementary structures and their subsequent assembly into the folding nucleus \cite{jchemphys2}. Such local elementary structures, being the elements which lead the folding process, have to be strongly stabilized at the very stages of the kinetics and, consequently, must display large values of the Eigenvector components. In the case of $\alpha$--spectrin SH3, the distal hairpin plays the role of local elementary structure and, interacting with the second strand of the RT loop, forms the folding nucleus (as discussed by Martinez\cite{serrano}). 

According to Fig. \ref{vec_md}(a), also the first strand of the RT loop (residues 10--15 (15--20)) play an importart role, if not for the folding and stability of the whole protein, at least for the stabilization of its N--terminal region. 

\subsection{src SH3 domain}

The SH3 domain of src protein displays the same fold as the $\alpha$--spectrin domain discussed above, although their sequence homology is only 36\%. Also for src SH3 the lowest Eigenvalue is well separated from the others, being $\Delta\lambda=18.7 kJ/mol$ and the average separation $\overline{\lambda}=4.0 kJ/mol$ (see Fig.  \ref{ev_md}(c) ). The components of the corresponding Eigenvectors are displayed in Fig. \ref{vec_md}(b).

Baker and coworkers performed a mutation analysis on src SH3 domain \cite{Grantcharova98} concluding that mutating residues G34 (G40), W36 (W42), I50 (I56) and Y54 (Y50) to A causes a destabilization of the native state ranging from 5.8 to 16.7 kJ/mol. It is worth noting that these sites are located in the region of the protein where the introduction of mutations to A determines $\varphi$--values ranging from 0.7 to 0.9. Furthermore, they showed that residues D9 (D15), Y10 (Y16), S12 (S18) and L18 (L24) are important for the protein stability, in that their mutations to  A causes destibilizations ranging between 4.1 and 17.1 kJ/mol. 
Also in this case there is a good agreement between the sites playing a major role in the stability and the kinetics of the protein and the peaks in the components of the Eigenvector (residues 10--20, 35--37, 49--51 and 54, cf. Fig. \ref{vec_md}(b)). Only site 23 and 34 are found in experiments to be important for the stability of the protein but display a small component in the Eigenvector. This could be due to the fact that these sites are occupied by glycines which, lacking the sidechain, may display a lower number of interactions with the flanking amino acids in MD simulations. 

The comparison of the Eigenvectors associated with $\alpha$--spectrin and src SH3 domains shows a remarkable similarity, although the sequences are rather different (only 36\% homology). This fact suggests a stronger evolutionary relationship between the two protein than what the mere comparison of sequences would indicate. In other words, $\alpha$--spectrin and src SH3 domains have diverged enough to change 74\% of their amino acids, but not enough to mutate their topologies and energetic pattern, and consequently their folding mechanims, with respect to each other. 

This aspect is reflected by the chemical nature of the residues at the "hot sites" sites of the two proteins. Both SH3 domains display peaks corresponding to acidic (D9) and hydrophobic (V18 for $\alpha$--spectrin SH3 and L18 for src SH3) residues in the N-terminal region, and sterically demanding hydrophobic (W36 for $\alpha$--spectrin SH3, W37 for src SH3) together with aliphatic (V48 for $\alpha$--spectrin SH3, I50 for src SH3) at the C--terminal. Moreover, these hot sites are part of a defined, mostly hydrophobic core coincident with the mechanical nucleus defined by Martinez \cite{serrano}, formed during the folding process of the two domains studied. Based on their results, those authors showed that passing through the transition state barrier  in SH3 domains requires the formation of a defined structure with little conformational variability in well defined regions, such as the one identified here. 
It is worth noting that the regions reach in hot sites, involving residues 34--39 and 46--49 are the peptide ligand binding regions, suggesting that the conservation of the enegetic pattern also assumes a functional role: well defined interactions among certain residues can be evolved in paralled to accomplish multiple tasks such as stabilization, binding etc. This aspects highlight the importance of the topology of the folded state in determining the role single residues have to play with regards to stability and function.
We can also conclude that the analysis of the energetic patterns of proteins sharing the same fold (topology) could be used to assess their evolutionary relationships in a way which is more efficient than the comparison of the sequences, because proteins have some degree of freedom of mutating their amino acids (even the key amino acids) for functional reasons, but the energetic pattern has to remain the same.

\subsection{Proteins G and L}

Proteins G and L are particularly suitable for our analysis because they have been extensively studied by Baker and coworkers \cite{Kim00,McCallister00}, who measured the effect of mutations in every site of the two proteins. 

\begin{figure}
\centerline{\psfig{file=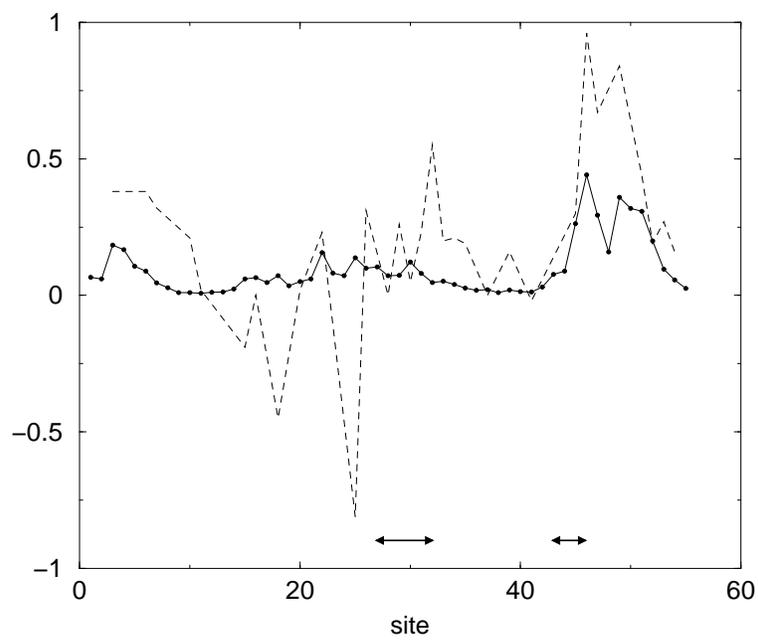,width=10cm,angle=-90}}
\caption{The $\varphi$--values of protein G (solid curve) and the corresponding components of the Eigenvector corresponding to the lowest Eigenvalue (dashed curve, same data as in Fig. 3(c)).}
\label{G}
\end{figure}

The $\varphi$--values of protein G are displayed in Fig. \ref{G} and display a high peak in the interval 45--52 and two small peaks around residues 3 and 22. Moreover, residues D22, A26 and F30 are important for the stability of the native state, their mutation to A causes a decrease of the stabilization free energy of 7.3, 12.3 and 5.9 kJ/mol, respectively. 

The spectrum of Eigenvalues for protein G, as in the case of SH3 domains, displays a gap between the lowest and the next--to--lowest level $\Delta\lambda=15.7$ kJ/mol, much larger than the average distance between the others $\overline{\lambda}=4.1$. The components of the associated Eigenvector are also displayed in Fig. \ref{G} (dashed curve). The highest peak in this plot matches exactly with the peak in the $\varphi$--values at sites 45--52, corresponding to the second hairpin whose formation is critical to the folding mechanism \cite{BakerL}. Moreover, also the other minor peaks in the Eigenvector components at sites 3--7, D22, A26 and V29 correspond to the others sites relevant to the stability and to the kinetics (cf. Fig. \ref{G}).

The residues relevant for stabilization are involved in the definition of the binding region of the IgG binding domain from protein G \cite{Hellinga}: mutations of residues in the region 27--32 and 43--46 causes major variations in the binding constant of human immunoglobulin Fc fragment. In this case too, the analysis of the protein energetics allows to undercover the convergence between the determinants of the stabilization and of the function of protein G. 

\begin{figure}
\centerline{\psfig{file=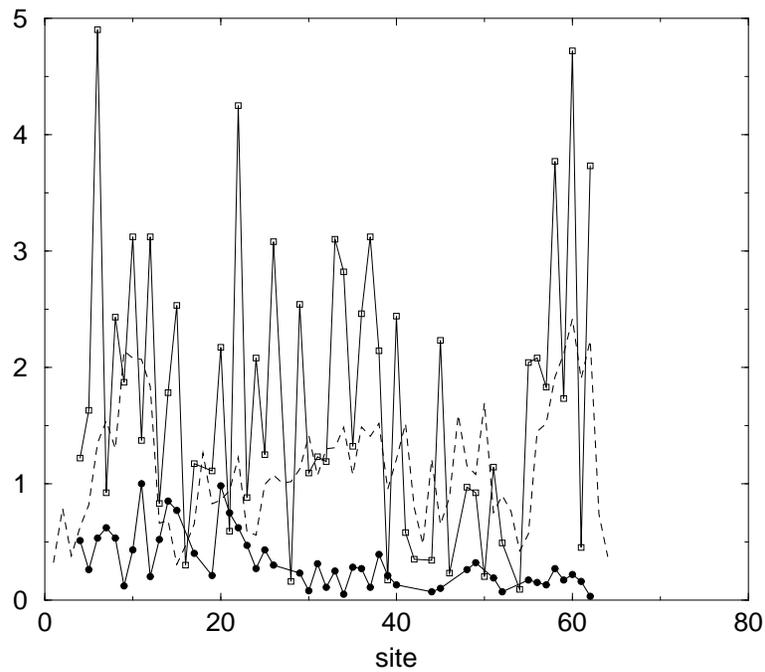,width=10cm,angle=-90}}
\caption{The $\varphi$--values (solid thick curve with filled circles) and stability of the native state (solid thin curve with empty squares) of protein L  and the corresponding components of the Eigenvector corresponding to the lowest Eigenvalue, multiplied by 10 for sake of readibility (dashed curve, same data as in Fig. 3(d)).}
\label{L}
\end{figure}

Protein L, although sharing the same fold with protein G, has a different folding mechanism. According to Baker \cite{BakerL} the critical step is the formation of the first hairpin (residues 1--22), the corresponding residues displaying $\varphi$--values higher, in average, than the rest of the protein (cf. Fig. \ref{L}, thick solid curve). The residues responsible for the overall stability of the protein are, aside from the first hairpin, F22, E27, T30, Y34, A35, A37--T39 and K54--L58 (cf. thin solid line in Fig. \ref{L}).

The value of $\Delta\lambda$ arising from the simulations is 17.84 kJ/mol, to be compared with $\overline{\lambda}=3.88$ kJ/mol [controllare] (see Fig. \ref{ev_md}(d)), so that also in the case of proten L it is meaningful to study the components of Eigenvector associated with the lowest Eigenvalue. The peaks in these components match well the sites with large $\varphi$--values and with large contribution to the stability of the native state (cf. Fig. \ref{L}). The values assumed by these components are, anyway, more uniformly distributed than the others proteins discussed sofar. 

The plot of the Eigenvector components for protein G and protein L display some common feature, such as the pattern of three intervals where they assume large values, corresponding to the first hairpin, to the helix and to the second hairpin, respectively. The relative heigth of these peaks is however different. In the case of protein G the peak associated with the second hairpin is quite larger than the others, consistently with the importance of this hairpin in the folding process and in the protein stabilization. The plot referring to protein L is more symmetric: in this case, the first  \(\beta \)-hairpin plays a relatively more important role in the stabilization of the native state, the second one being energetically disfavored because of the high backbone torsional strain due to three consecutive residues in the \(\beta \)-turn having unfavorable positive \( \Phi \) angles.

\subsection{Chymotrypsin Inhibitor 2}

Chymotrypsin Inhibitor 2 (CI2) is a 64-residue polypeptide inhibitor of serine proteases \cite{Bycroft91}. CI2 is a single module of structure: the interatomic interactions are quite uniform over the structure, and they do not segregate into regions that make more tertiary interactions within themselves than they do with neighboring atoms. As such, CI2 can be considered a basic folding unit or foldon \cite{Panchenko96}. This is reflected by the fact that most $\varphi$--values of this protein are fractionary \cite{ci2}.

This peculiarity of CI2 is underlined both by the fact that the difference between the first two lowest eigenvalues is $\Delta\lambda=13.2$ kJ/mol, the average spacing being  $\overline{\lambda}=3.9$, and by the absence of well defined isolated peaks in the components of the Eigenvector associated to the first eigenvalue. 

Nonetheless, a careful inspection of Fig. \ref{vec_md}(e) shows that the stabilizing regions of the molecule are located around residues 10--20 and 40--50, corresponding to the helix of the N-terminal region making contacts with residues located on a \(\beta \)-sheet at the opposite end of the polypeptide. Mutations in these regions cause destabilization of the folded state between 4.0 and 20.6 kJ/mol. The long range interactions between these regions are those required to stabilize the helix in peptide fragments of CI2, as shown by Fersht and coworkers in a study on the stability of the secondary structure of peptides isolated from CI2 \cite{fershtA}. Moreover, residues around position 16, 49 and 57 consititute a nucleation site, whose disruption leads to loss of stability and fast unfolding of the protein at high temperature conditions \cite{Kazmirski01,Gay95,Neira97}.

\section{Discussion}

\begin{figure}
\centerline{\psfig{file=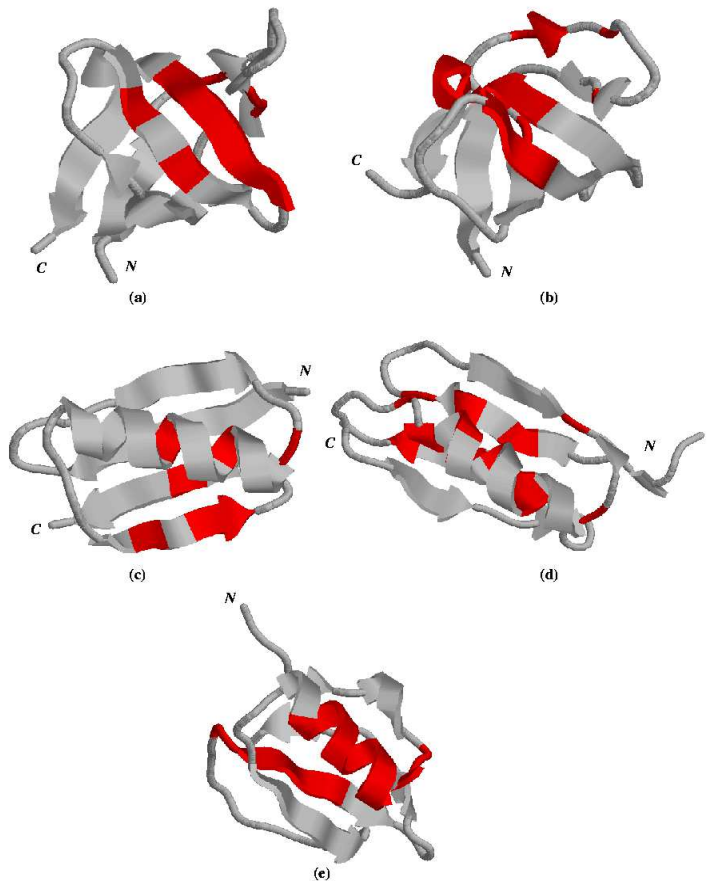,width=10cm}}
\caption{The hot sites mapped on the three dimensional structure for (a) \( \alpha  \)-spectrin SH3 domain, (b)  src SH3 domain, (c) protein G, (d) protein L, and (e) CI2.  The hot sites are part of the folding nucleus.}
\label{hot_spot}
\end{figure}

By simulating the equilibrium fluctuations of a protein around its native conformation through molecular dynamics simulations it is possible to investigate the energetic pattern (i.e., interaction energy, not free energy) which defines the native state and contributes to make it a minimum in the free energy landscape. A feature common to all proteins which emerges from such calculations is that, in each protein, few residues build clusters of strong interactions, surrounded by the other, weakly interacting residues (see Fig. \ref{hot_spot}). This fact can be mathematically assessed noting that the lowest Eigenvalue of the residue--residue interaction matrix is well separated from the others, and the clusters can be detected searching for the peaks in the components of the associated Eigenvector.

The presence of a strongly interacting nucleus is not unexpected and has been observed in minimal lattice models of proteins \cite{nucleus,jchemphys2}. This nucleus not only gives account for the all--or--none character of the folding transition \cite{sh1} and for its remarkable tolerance to point mutations \cite{jchemphys1}, but is also responsible for the fast folding of the protein. In fact, 
the hierarchical assembly of local elementary structures, composed of strongly interacting residues lying close along the chain, has been shown to drive the protein to the native state in the case of simple models \cite{jchemphys2} (see Introduction). This structure is necessarily associated with a cluster of strong interactions composed of those which stabilize the local elementary structure and those which stabilize the nucleus, keeping together the local elementary structures.

The analysis of the components of the Eigenvector associated with the lowest Eigenvalue of the native--state interaction matrix is helpful in determining such a cluster. Consequently, the sites individuated by this analysis play a major role in the kinetics and stability of the protein (Fig. \ref{hot_spot}), as shown by the good agreement between the plot of the Eigenvector components and the plots displaying $\varphi$--values and change in stability upon mutations. From this analysis it is not possible to tell which of the sites indicated by the Eigenvector are connected with stability, which with kinetics and which with both of them. The Eigenvector contain a superposition of the two informations, and only an analysis of the entropy of the system can separate them. From this point of view, the present analysis consists in an approximation of the free energy of the native state where the entropic term is neglected, and consequently is quite obvious that the results it can give are partial.

It is however interesting that the energetics of the native state contains informations about the folding kinetics. This could sound strange, but is not as strange as stating that the sequence alone contain all the information about the folding process \cite{anfinsen}. In fact, if the hierarchical picture of folding arising from minimal models is correct, it is possible from the energetic pattern and from the entropics of the native state to find the local elementary structures and the nucleus. Since the information about the folding mechanism has to be contained in the sequence of amino acids \cite{anfinsen}, and the sequence determines the chemical nature of the protein and, consequently, the interaction pattern, it is the energetics where we have to focus our attention to localize the local elementary structures. Of course, the energetic pattern of the native state alone cannot contain informations about metastable states and kinetic traps, something which can affect the kinetics to a large extent, expecially in the case of longer proteins.

To be noted that the average spacing between the peaks in the Eigenvector component of all the proteins we analyzed is $\approx 19$ residues. This is in accordance with the findings of Berezovsky and Trifonov \cite{trifonov,trifonov2}, who analyze the size of closed loops of the backbone of a large number of proteins and observe an average ranging from 25 to 30 residues, a size which is optimal from the point of view of a fast search in conformational space. We suggest that the loops found by Berezovsky and Trifonov in ref. \cite{trifonov,trifonov2} could be related to the local elementary structures discussed above \cite{jchemphys2}.

From the practical point of view, the analysis of the components of the Eigenvector associated with the lowest Eigenvalue of the native--state interaction matrix can be  helpful in determining the positions or the regions where mutations have to be introduced in order to change the stability, and connected with it, the dynamical features of the folded protein. Once the hot sites have been identified, saturation mutagenesis experiments can be focused on that region to modify the properties of the protein under study. In many cases \cite{Arnold98,Liebeton00} "hot sites" have been identified through several cycles of error--prone mutagenesis and analysis, which relies on the random introduction of mutations in different parts of the protein, and subsequent saturation mutagenesis on the above defined hot sites. This procedure, despite being quite efficient and successfull, is not based on a rational approach, and might be time consuming. The straightforward application of realistic MD simulations followed by the analysis of the eigenvectors of the energy interaction matrix can be a useful help in rational protein design.

Two caveats to the method presented here needs to be highlighted. First, the molecular dynamics simulations with explicit water we employ produce an interaction matrix which accounts for the Van der Waals energy, for the electrostatic energy and for the hydrogen bonds energy, but misses the hydrophobic interaction. This is, in fact, an effective interaction which arises from the averaging of the solvent degrees of freedom and does not appear explicitely in the potential function of the system. The fact that we obtain good agreement between these model calculations and the experimental results, suggests that the hydrophobic interaction does not play a major role in determining the local elementary structures, but only in the generic collapse of the chain into a globule (which is not considered here). Sensitive differences in the results for both eigenvectors and eigenvalues arise when a cutoff scheme is used to calculate non bonded electrostatic interactions. In general there is little if not no agreement between the calculated energy stabilizing residues and the experimental results; a fact that is not unexpected if one considers that cutoff methods truncate the interactions of charged atoms in space, neglecting interactions through space which might play a role in the formation of the stabilizing nucleus.
Another problem is the choice of the time duration of the simulations. They have to be long enough to average the uncertaninties in the structure connected with the experimental determination of the native conformation and to be equilibrated. Running simulations where multiple folding-unfolding events could be observed would give access to the entropic term which has been neglected in the analysis, but reaching the timescales required for that purpose is currently out of reach even for small proteins. We have nonetheless performed simulations ranging from 2 to 15 ns, observing no qualitative differences between the different cases belonging to this range.

\section{Conclusions}

We have shown that it is possible to identify the sites of a protein which most contribute to the stability and to the accessibility of the native state by mean of a simple energetic analysis of molecular dynamics simulations, simulations that can be performed even for large proteins 
just with the help of a PC. This allows not only to gain insight into the folding pathways of a given protein, but also to make more effective mutagenesis experiments, by focusing only on a selected number of sites.

\section{Matherials and methods}

The starting structure for the five all--atom md simulations were taken from the protein data bank: 1SHG.pdb for $\alpha$ spectrin SH3 domain, 1FMK.pdb for src SH3 domain,1PGB.pdb for protein G, 2PTL.pdb for Protein L and 2CI2.pdb for chymotrypsin inhibitor 2.
The proteins were protonated to give a zwitterionic form (with N-terminal
NH\( _{3}^{+} \) and C-terminal COO\( ^{-} \) groups) with the carboxyl NH side chain groups in their charged states. The total charges on the proteins resulted \( +1 \) for $\alpha$ spectrin SH3 domain, \( -4 \) for src SH3 domain, Protein G and Protein L and \( -1 \) for CI2.
The proteins were solvated with water in a octahedral box large enough
to contain \( 1.2\, nm \) of solvent around the peptide. The simple point
charge (SPC) water model was used~\cite{Berendsen87} was used to solvate each protein in the simulation box. 
In all simulations, suitably charged counterions were used  to yield an electrically neutral system. Each system was subsequently energy minimized with a steepest descent method for \( 1000 \) steps. 
The calculation of electrostatic forces utilized the PME implementation of the Ewald summation method.
The LINCS algorithm ~\cite{Hess97} was used to constrain all bond lengths.
For the water molecules the SETTLE algorithm ~\cite{Miyamoto92} was used. A
dielectric permittivity, \( \epsilon =1 \), and a time step of \( 2 \)~fs
were used. All atoms were given an initial velocity obtained from a Maxwellian distribution at the desired initial temperature of 300K. The density of the system was adjusted performing the first equilibration runs at NPT condition by weak coupling to a bath of constant pressure (P\( _{0}=1 \) bar, coupling time \( \tau _{P}=0.5 \) ps) ~\cite{Berendsen84}.  
In all simulations the temperature was maintained close to the intended values by weak coupling to an external temperature bath ~\cite{Berendsen84} with a coupling constant of \( 0.1 \)~ps. The peptide and the rest of the system
were coupled separately to the temperature bath. Each of the five MD simulations was extended to 10 ns. 
All simulations and  analysis were carried out using the GROMACS package (version
3.0 ~\cite{GROMACS}), using the GROMOS96 43A1 force
field~\cite{vanGunsteren98}. 
All calculations were performed on clusters of PCs, with Linux operating
system.

\newpage    
\bibliographystyle{jmb}
\bibliography{/usr/people/giorgio/PAPERS/BIB/bib}

\newpage
\begin{table}
\begin{tabular}{|c|c|c|}
\hline
& $\Delta\lambda$ kJ/mol &  $\overline{\lambda}$ kJ/mol\\\hline
lattice protein & 108.4 & 18.8 \\
$\alpha$-spectrin SH3 & 20.9 & 4.2 \\
src SH3 & 18.7 & 4.0 \\
protein G & 15.7 & 4.1 \\
protein L & 17.8 & 3.9 \\
CI2 & 13.2 & 3.9 \\\hline
\end{tabular}
\caption{The difference $\Delta\lambda$ between the first two Eigenvalues and the average spacing $\overline{\lambda}$ between the other Eigenvalues.}
\end{table}

\begin{table}
\begin{tabular}{|c|c|c|}
\hline
& $\Delta\Delta G_{N-U} [kJ/mol]$ & $\varphi_{\ddagger-U}$ \\\hline
L3 & 3.38 & 0.24 \\
17 & & 0.45 \\
V18 & 7.53 & \\
T19 & 4.18 & 0.17\\
L28 & 5.18 & -0.3\\
31 & & 0.3\\ 
K38 & 2.50 & 0.4\\
V39 & 12.95 & 0.55\\
43 & & 1\\
V48 & 8.98 & \\\hline
\end{tabular}
\caption{Effect of point mutations on the amino acids of $\alpha$--spectrin SH3 domain, as found by Martinez, Pisabarro and Serrano \protect\cite{serrano}. Note that the numeration of amino acids has been changed in such a way that the first is numbered with 1 (instead of 5).}
\end{table}

\newpage

\end{document}